# Thermally and Molecularly Stimulated Relaxation of Hot Phonons in Suspended Carbon Nanotubes

David Mann,[1] Eric Pop,[1,2] Jien Cao,[1] Qian Wang,[1] Kenneth Goodson[2] and Hongjie Dai[1,*]

[1] *Department of Chemistry and Laboratory for Advanced Materials, Stanford University, Stanford, CA 94305, USA*

[2] *Department of Mechanical Engineering and Thermal Sciences, Stanford University, Stanford, CA 94305, USA*

The high-bias electrical transport properties of suspended metallic single-walled carbon nanotubes (SWNTs) are investigated at various temperatures in vacuum, in various gases and when coated with molecular solids. It is revealed that non-equilibrium optical phonon effects in suspended nanotubes decrease as the ambient temperature increases. Gas molecules surrounding suspended SWNTs assist the relaxation of hot phonons and afford enhanced current flow along nanotubes. Molecular solids of carbon dioxide frozen onto suspended SWNTs quench the non-equilibrium phonon effect. The discovery of strong environmental effects on high current transport in nanotubes is important to high performance nanoelectronics applications of 1D nanowires in general.

* E-mail: hdai@stanford.edu.

The effect of environment on the electrical transport properties of small structures is an interesting topic with implications to a wide range of applications, from high performance nanoscale electronic devices and interconnects to nanosensors. Thus far, it has been largely unexplored how extrinsic factors affect high field electron transport in 1D nanostructures, and how to exploit them to tune and manipulate the high-current carrying abilities of 1D materials. As an extreme example, it has been shown recently[1] that freely suspended SWNTs *in vacuum* (in an isolated environment) display negative differential conductance (NDC) and drastically reduced current levels compared to nanotubes lying on solid substrates,[2,3] caused by substantial self-heating and scattering by non-equilibrium optical phonons (OPs) in the suspended isolated SWNTs.[1] Here, we present the first systematic investigation of how various environmental factors affect high-field electron transport, phonon scattering and phonon relaxation in SWNTs. These factors include ambient temperature, gas molecule pressure and type surrounding nanotubes and molecular-solid condensates.

Suspended ($L \sim 1$-3 $\mu$m long, $d \sim 1.8$-3.8 nm diameter) metallic SWNTs were obtained by CVD growth across pre-formed trenches and Pt electrodes (Fig. 1) as described previously.[4,5] The devices were characterized by scanning electron microscopy (SEM) to obtain nanotube length information. Electrical measurements were carried out in a variable-temperature environmental probe station in vacuum ($\sim 10^{-6}$ torr) or various gases between temperatures $T_0 = 250$-400K, or at $T_0 \sim 50$K with condensed $CO_2$. It is important to note that prior to measurement, each sample was first baked at 400K in high vacuum ($< 10^{-6}$ Torr), then left in vacuum overnight to desorb gases from the contacts and minimize systematic variations in sample treatment and history.

We measured large numbers of individual suspended SWNTs in vacuum at various $T_0$ (defined as the metal electrode temperature, (Fig. 1c) between 250K and 400K and always observed their hallmark negative differential conductance (NDC) behavior at high biases (Fig. 2a).[1] Interestingly, the *I-V* curves taken at different $T_0$ tend to converge in the high bias regime, and NDC appears less pronounced at higher $T_0$ (Fig. 2a). To understand the *I-V* data, we adopt and extend a recently introduced electro-thermal model for suspended SWNTs.[1] Briefly, the resistance of a SWNT under self-heating can be written as

$$R(V,T) = R_c + \frac{h}{4q^2} \frac{L + \lambda_{eff}(V,T)}{\lambda_{eff}(V,T)}, \qquad (1)$$

where $R_c$ is the contact resistance and $\lambda_{eff} = \left(1/\lambda_{ac} + 1/\lambda_{op,ems} + 1/\lambda_{op,abs}\right)^{-1}$ is the bias and temperature dependent electron mean free path (MFP), including scattering with AC phonons, and OP emission and absorption.[1] The current is computed as *I=V/R*. The lattice temperature distribution along the SWNT is obtained by solving the heat conduction equation with $T=T_0$ boundary conditions at both ends of the tube[6]

$$A\nabla(\kappa_{th}\nabla T) + p' - g(T - T_0) = 0, \qquad (2)$$

where $\kappa_{th}$ is the SWNT thermal conductivity, $p' = I^2(R-R_c)/L$ is the Joule heating per unit length, $A=\pi db$ is the cross-sectional area ($b\sim0.34$ nm: tube wall thickness) and $g$ is the net heat loss by radiation and heat conduction per unit length. With $\kappa_{th}$ vs. $T$ as fitting parameters, we solve Eqs. (1) and (2) iteratively until the temperature profile at each point along the SWNT converges within 0.1 K, while obtaining the best current fit with experimental data, at every bias. This temperature profile corresponds to AC phonons ($T_{ac}$), which are the main heat carriers in the temperature range considered.[7] We capture the non-equilibrium OP effects in suspended SWNTs with an effective OP temperature[1]

$$T_{op} = T_{ac} + \alpha\left(T_{ac} - T_0\right) \qquad (3)$$

where the non-equilibrium phonon coefficient $\alpha > 0$. The physical picture is that OP phonons emitted by hot electrons relax by decay into AC modes ($T_{op}$ to $T_{ac}$) that are subsequently carried out of the tube through the contacts ($T_{ac}$ to $T_0$). OP decay by direct propagation out of the tube is not appreciable due to their low group velocity as compared to AC modes. With $\kappa_{th}$ and $\alpha$ as fitting parameters, eqs. (1)-(3) are used to calculate $T_{ac}$ and $T_{op}$ along the tube, the respective AC and OP scattering MFPs, and consequently the *I-V* characteristics of the nanotube to fit the experimental data.

In vacuum where $g \sim 0$, Joule heating in the tube dissipates along its length to the contacts, resulting in a parabolic temperature profile along the SWNT. For the $\kappa_{th}(T) \sim 1/T$ dependence (consistent with umklapp phonon scattering at high temperatures),[1,8] our model enables good reproduction of the experimental *I-V* curves in the high-bias region (Fig. 2a) and numerical extraction of $\alpha$ (Fig. 2b). The non-equilibrium coefficient is found to behave as $\alpha \sim 2.3(300/T_0)^m$ with $m \sim 1.5$ (Fig. 2b), suggesting reduced non-equilibrium OPs when the suspended SWNT is at a higher ambient $T_0$.

We give a theoretical account for the non-equilibrium phonon coefficient $\alpha$ as follows. The average AC temperature can be written as $T_{ac} = T_0 + P\mathcal{R}_{th}$ and the average OP temperature as $T_{op} = T_{ac} + P\mathcal{R}_{op}$, where $\mathcal{R}_{op}$ and $\mathcal{R}_{th}$ are the *thermal* resistances for OP decay into AC phonons and for AC heat conduction along the tube, respectively. The Joule power $P$ is dissipated first to the OP modes which then decay into AC modes, a sensible assumption for suspended SWNTs under high bias, when transport is limited by high-energy ($\hbar\omega_{op} \sim 0.18$ eV) OP scattering.[1] Consequently the OP non-equilibrium coefficient $\alpha$ can be written as

$$\alpha = \frac{\mathcal{R}_{op}}{\mathcal{R}_{th}} \sim \frac{\tau_{op}/C_{op}}{L/(12A\kappa_{th})} \sim \left(\frac{12A}{L}\right)\left(\frac{\tau_{op}\kappa_{th}}{C_{op}}\right). \quad (4)$$

The OP lifetime $\tau_{op} \sim 1/T$ scales approximately as the phonon occupation[8] and is consistent with previous observations of OP Raman linewidths $\propto T$.[8, 9] The thermal conductivity $\kappa_{th} \sim T$ at low $T$ (in 1-D) and $1/T$ at high $T$,[1, 8] and the OP heat capacity $C_{op} \sim T$ at low $T$ and approaches a constant at high $T$.[8] These lead to an expected temperature dependence of $\alpha \sim 1/T$ at low $T$ and $1/T^2$ at high $T$ (see trend lines in Fig. 2b). Our extracted $\alpha \sim 1/T_0^{1.5}$ falls within the theoretically expected trends (Fig. 2b). These results represent the first observation of ambient temperature stimulated reduction of non-equilibrium OPs and its effects on high-bias transport in SWNTs. The reduced non-equilibrium effect at higher $T_0$ is owed to stimulated OP to AC decay at higher temperatures and the $\kappa_{th} \sim 1/T$ AC thermal conductivity (Eq. 4).

Next, we investigate the effects of molecules to non-equilibrium phonons in suspended SWNTs (at 300K). We have consistently observed increases in the high-bias currents of nanotubes in various gases and pressures when compared to in vacuum (Fig. 3a, 3b). We have considered gas molecules' effect on conduction heat loss (or heat sinking) ($g$ factor in Eq. 2) and non-equilibrium OPs ($\alpha$) in suspended SWNTs. An upper limit of $g \sim (nv/4)(3k_B/2)(\pi d) = 0.428$ mWK⁻¹m⁻¹ is estimated[10] for a nanotube in 1 atm of $N_2$, where $n$ is molecule density and $v$ is the average molecular velocity (dependent on gas type via molecular mass). This is much smaller than $g \sim 100$ mWK$^{-1}$m$^{-1}$ estimated for tubes lying on solid $SiO_2$ substrates,[6] indicating little heat sinking by the gaseous molecules around the nanotube. The combined SWNT heat loss to the gas ambient and to radiation is estimated to be < 2 % of the power dissipation at all relevant temperatures. By including the $N_2$ $g$ factor in Eq. 2 and using the $\alpha$ coefficient in vacuum, our modeling fails to reproduce the high-bias current enhancement by $N_2$ seen experimentally. On the other hand, when allowing the variation of $\alpha$ from its value in vacuum, our calculations successfully reproduce the *I-V* curves of suspended SWNTs in various $N_2$ pressures (Fig. 3a). A systematic decrease in $\alpha$ is found as the gas pressure increases, indicating reduced non-equilibrium OPs. This is an interesting result that reveals gas molecules stimulating the relaxation of OPs (reduced $\alpha$ ) in suspended SWNTs and thus enhancing the high-bias current carrying capability of nanotubes. We note that the *low bias* resistance of the suspended tubes is *not* affected by the presence of gases studied here (Fig.3a,3b), indicating negligible gas chemical gating effect[11] to the nanotubes. The hot phonon relaxation effect by molecules also differs from the recently reported molecular indentation effect.[15]

By varying the type of gas molecules surrounding suspended SWNTs (at 300K, in 1 atm. pressure), we find that polyatomic gases are more effective than monatomic gases in relaxing hot OPs in suspended SWNTs, and the degree of relaxation (reduction in $\alpha$) increases with the number of atoms in the gas molecule. (Fig. 3b,3c)  Although He has a higher thermal conductivity than Ar, it affords slightly lower current enhancement in nanotubes than Ar, again suggesting that heat conduction is not primarily responsible for the observed gas effect. Based on these results, we propose that the mechanism of OP relaxation by molecules is the coupling of phonon modes in the nanotube with various degrees of motion of the molecules including the vibrational modes. Notably, coupling between the vibrational modes of physisorbed molecules and surface OPs of solids have been previously documented.[8, 12, 13] For nanotubes, infrared spectroscopy has clearly revealed vibrational frequency shift of $CO_2$ molecules when physisorbed on SWNT surfaces due to molecule-nanotube interaction.[14] Our current work suggests that the physisorbed gas molecules on SWNTs are also capable of affecting and relaxing energetic OP modes in nanotubes.

For the un-reactive gases used, molecule-SWNT interaction is expected to be van der Waals (binding energies ~30-100 meV)[14] leading to short absorption resonance times on nanotubes at 300K. If molecules are 'fixed' on a nanotube surface, a high degree of SWNT-molecule vibrational coupling can be expected. To test this hypothesis, we cooled our suspended SWNTs to 50K and encased the SWNT in solid $CO_2$ (dry ice) by leaking $CO_2$ gas into the cryogenic probe-station. Upon solid $CO_2$ condensation on the nanotubes, drastic effects are observed in the I-V characteristic of nanotubes including elimination of NDC and enhanced currents, approaching 20μA (Fig. 4). Similar to SWNTs lying on solid $SiO_2$ substrates,[1] efficient phonon and thermal coupling exist between SWNT/solid $CO_2$ affording little self-heating and non-equilibrium optical phonon effects (i.e., large $g$ and $\alpha \sim 0$). Importantly, the condensation of $CO_2$ caused no adverse effects on the SWNT and was highly reversible upon repeated condensation and evaporation of $CO_2$.

We have systematically revealed thermal and molecular relaxations of non-equilibrium optical phonons in suspended SWNTs and their effects on SWNT high-bias electrical transport properties.  High ambient temperature reduces non-equilibrium optical phonons by stimulated decay to acoustic phonons.  Hot phonons in nanotubes can be coupled and relaxed by surrounding molecular gases and solids.  To our knowledge, this is the first time that phonon coupling with adsorbed molecules on surfaces is found to affect high field electrical transport properties in any bulk or nanoscale solid. For 1D nanowires or 0D dots with ultra-high surface areas relative to volume, adsorbed surface species can profoundly affect how current flow in these structures, which should be understood, avoided or exploited.

**Acknowledgement.**  We thank Prof. W. Harrison for insights.  Project supported by MARCO MSD.

**Figure Captions:**

**Figure 1.** Suspended nanotube devices. (a) SEM image of a device showing a trench and source (S) and drain (D) electrode structures. (b) Close-up SEM (1kV acceleration voltage) of a SWNT suspended over the trench. (c) Schematic drawing of the device. Metallic and quasi-metallic nanotubes under a high negative gate voltage (in fully "on" state) are used for all measurements in this work.

**Figure 2**. Electrical transport characteristics of a suspended nanotube in vacuum at various ambient temperatures. (a) *I-V* characteristics of a SWNT with $L \sim 2.6$ μm, $d \sim 1.7$ nm and $R_c \sim 8$ kΩ at four ambient $T_0$: experimental data (symbols) and model calculations (lines) based on SWNT thermal conductivity $\kappa_{th}(T) \sim 3400(300/T)$ Wm$^{-1}$K$^{-1}$. Data fitting at high bias (> 0.3 V) allows the extraction of the OP non-equilibrium coefficient $\alpha$ at each $T_0$. (b) Extracted OP non-equilibrium coefficient $\alpha$ (symbols) from (a) and trend lines as a function of contact/ambient temperature $T_0$. Decreasing $\alpha$ indicates reduced non-equilibrium OP effects at higher ambient temperatures.

**Figure 3**. Electrical transport characteristics of suspended nanotubes in vacuum and various gases. (a) Experimental *I-V* data (symbols) and model calculations (lines) of a nanotube device ($L \sim 2.1$ μm, $d \sim 2.4$ nm and $R_c \sim 4$ kΩ) in several $N_2$ pressures. The gas pressure-dependent data can be reproduced by model (Eq. 1-3) calculations by varying $\alpha$ approximately linearly with pressure between 2.3 and 1.3. (b) *I-V* curves of a device ($L \sim 2.3$ μm, $d \sim 3.8$ nm and $R_c \sim 13$ kΩ) in vacuum, Ar, $N_2$ and $C_2H_4$ at 1 atm. pressure respectively. The symbols are experimental data and the lines are model calculations revealing reduced non-equilibrium OP phonons from $\alpha \sim 2.4$ in vacuum to $\alpha_{Ar} \sim 1.8$, $\alpha_{N2} \sim 1.5$, and $\alpha_{C2H4} \sim 1.1$. No hysteresis in the high bias NDC region of the *I-V* curves is observed upon back and forth voltage scans. Small hysteresis exists at times beyond the current peak of the *I-V* curves in various gases but can be eliminated when the sample is baked (400K) and pumped on (10$^{-6}$ Torr) prior to measurement. (c) High-bias (1.5 V) current enhancement ($\Delta I$) by various gas ambient at 1 atm. pressure relative to vacuum vs. the number of atoms in the molecules investigated. The molecule induced current enhancement increases with the number of atoms in the molecule and is uncorrelated with the thermal conductivities of the gases, which follow: He > $CH_4$ > $O_2$ > $N_2$ > Ar> $C_2H_4$ > $CO_2$.

**Figure 4**. Electrical characteristics of a suspended nanotube in vacuum and when coated with solid dry ice respectively at 50K.  We cooled the device to 50K in vacuum, recorded *I-V*, leaked in $CO_2$ gas into the chamber to form a dry ice coating on the device surface (dry ice layer visible by CCD camera) and then recorded *I-V* with the solid $CO_2$ coating on the nanotube.

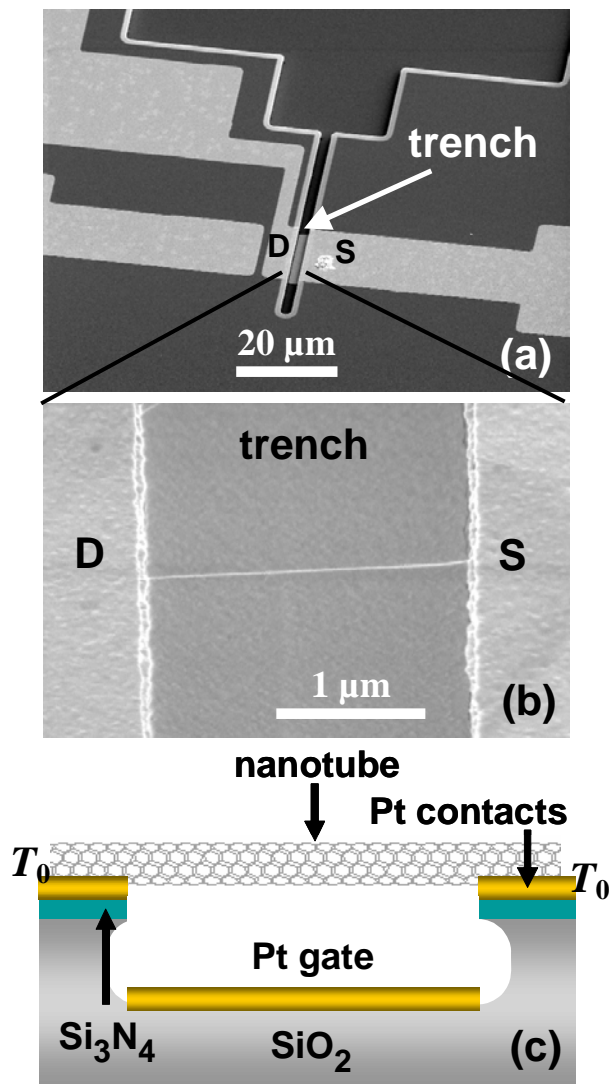

**Figure 1.**

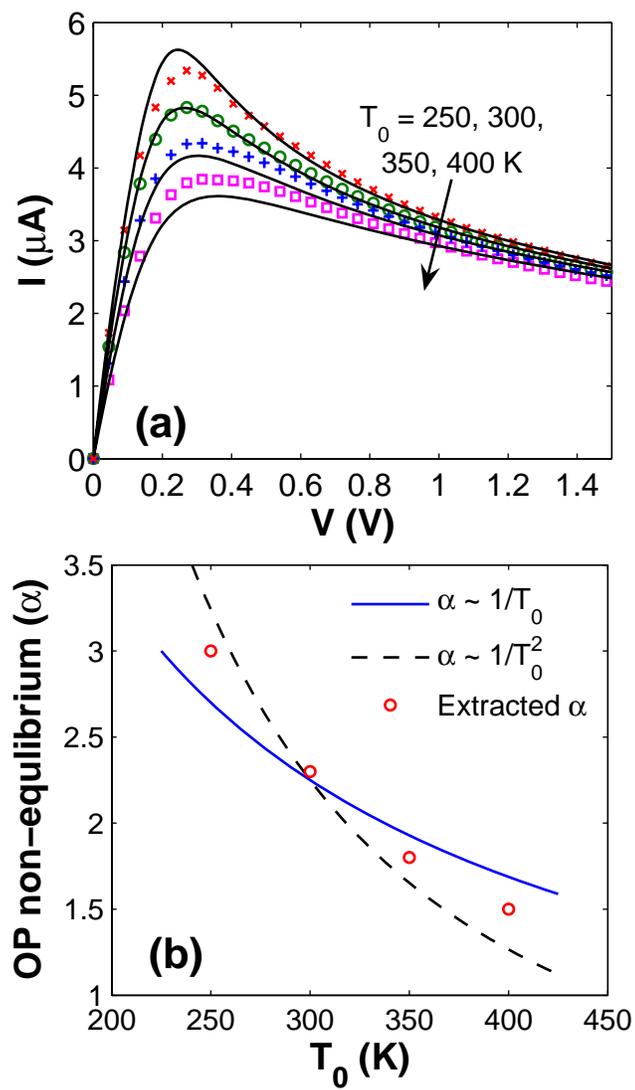

**Figure 2.**

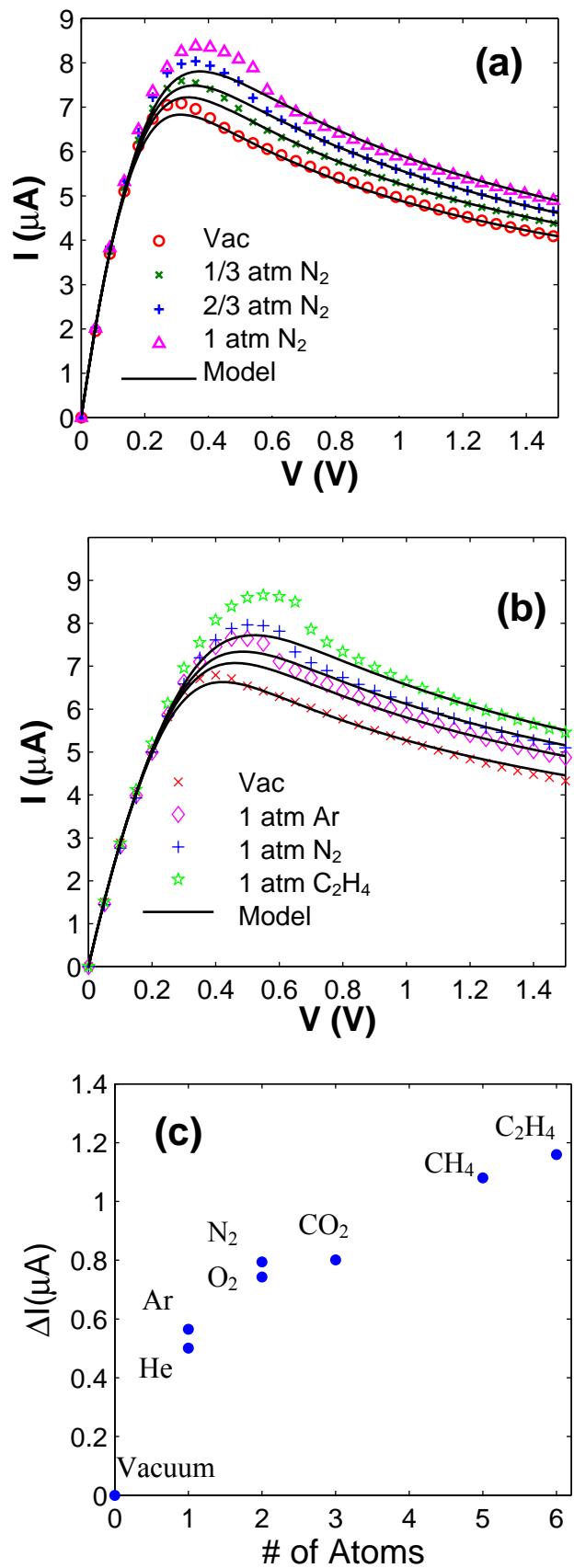

**Figure 3.**

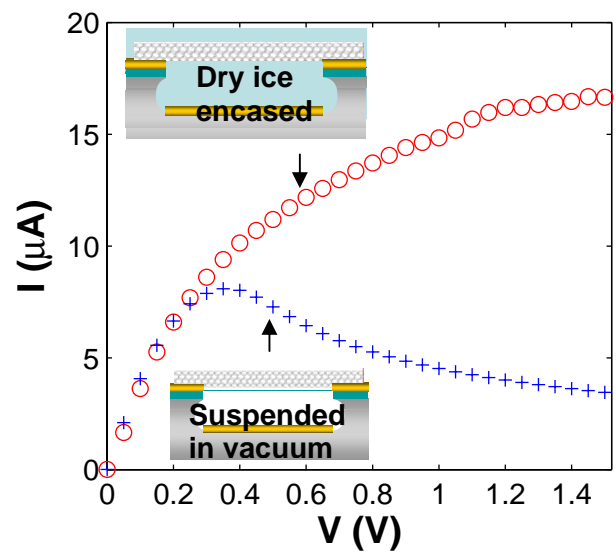